\documentclass[%
	10pt,
	english,
	showpacs,%
	amsmath,%
	amssymb,%
	aps,%
	twocolumn,%
	prl,
	notitlepage
]{revtex4-1}

\usepackage{graphicx}

\usepackage{ae}
\usepackage{xcolor}
\usepackage{multibib}

\usepackage[utf8]{inputenc}
\usepackage[T1]{fontenc}
\usepackage{pdfsync}
\usepackage[colorlinks=true,linkcolor=black,  
						anchorcolor=blue,  
						citecolor=blue, 
						filecolor=blue, 
						menucolor=blue, 
						urlcolor=blue]{hyperref}

\newdimen\capVert
\begin{document}

\title{Direct probing of the Mott crossover in the SU(\textit{N}) Fermi-Hubbard model}

\author{Christian Hofrichter}
\affiliation{Ludwig-Maximilians-Universit\"at, Schellingstra\ss{}e 4, 80799 M\"unchen, Germany}
\affiliation{Max-Planck-Institut f\"ur Quantenoptik, Hans-Kopfermann-Stra\ss{}e 1, 85748 Garching, Germany}
\author{Luis Riegger}
\affiliation{Ludwig-Maximilians-Universit\"at, Schellingstra\ss{}e 4, 80799 M\"unchen, Germany}
\affiliation{Max-Planck-Institut f\"ur Quantenoptik, Hans-Kopfermann-Stra\ss{}e 1, 85748 Garching, Germany}
\author{Francesco Scazza}
\affiliation{Ludwig-Maximilians-Universit\"at, Schellingstra\ss{}e 4, 80799 M\"unchen, Germany}
\affiliation{Max-Planck-Institut f\"ur Quantenoptik, Hans-Kopfermann-Stra\ss{}e 1, 85748 Garching, Germany}
\author{\mbox{Moritz H\"ofer}}
\affiliation{Ludwig-Maximilians-Universit\"at, Schellingstra\ss{}e 4, 80799 M\"unchen, Germany}
\affiliation{Max-Planck-Institut f\"ur Quantenoptik, Hans-Kopfermann-Stra\ss{}e 1, 85748 Garching, Germany}
\author{\mbox{Diogo Rio Fernandes}}
\affiliation{Ludwig-Maximilians-Universit\"at, Schellingstra\ss{}e 4, 80799 M\"unchen, Germany}
\affiliation{Max-Planck-Institut f\"ur Quantenoptik, Hans-Kopfermann-Stra\ss{}e 1, 85748 Garching, Germany}
\author{Immanuel Bloch}
\affiliation{Ludwig-Maximilians-Universit\"at, Schellingstra\ss{}e 4, 80799 M\"unchen, Germany}
\affiliation{Max-Planck-Institut f\"ur Quantenoptik, Hans-Kopfermann-Stra\ss{}e 1, 85748 Garching, Germany}
\author{Simon F\"olling}
\affiliation{Ludwig-Maximilians-Universit\"at, Schellingstra\ss{}e 4, 80799 M\"unchen, Germany}
\affiliation{Max-Planck-Institut f\"ur Quantenoptik, Hans-Kopfermann-Stra\ss{}e 1, 85748 Garching, Germany}
\email{simon.foelling@lmu.de}

\date{\today}

\maketitle


\textbf{
The Fermi-Hubbard model (FHM) is a cornerstone of modern condensed matter theory. 
Developed for interacting electrons in solids, which typically exhibit SU($2$) symmetry, it describes a wide range of phenomena, such as metal to insulator transitions  and magnetic order \cite{Imada1998}. 
Its generalized SU($N$)-symmetric form, originally applied to multi-orbital materials such as transition-metal oxides \cite{Tokura2000}, has recently attracted much interest owing to the availability of ultracold SU($N$)-symmetric atomic gases.
Here we report on a detailed experimental investigation of the SU($N$)-symmetric FHM using local probing of an atomic gas of ytterbium in an optical lattice to determine the equation of state through different interaction regimes.
We prepare a low-temperature SU($N$)-symmetric Mott insulator and characterize the Mott crossover, representing important steps towards probing predicted novel SU($N$)-magnetic phases\ \cite{cazalilla_ultracold_2014}.}


Although the Fermi-Hubbard Hamiltonian has been the object of a large number of studies in the past decades, reaching a complete understanding has remained an elusive task, even for the spin-1/2 case. For repulsive interactions, the SU(2) FHM is known to give rise to a paramagnetic Mott insulator, while anti-ferromagnetic order emerges below the Néel temperature. Moreover, the FHM is believed to capture the essential physics of \textit{d}-wave superfluidity in high-temperature superconductors \cite{Anderson1987,Imada1998}. 
The development of experimental implementations of the three-dimensional (3D) FHM with ultracold atoms has provided a new approach for advancing our understanding of strongly correlated fermions in lattices
\cite{Esslinger2010}.

The recent realization of degenerate gases of strontium and ytterbium in optical lattices allows to go beyond the conventional spin-1/2 FHM and to access the largely unexplored physics of the more general SU($N$) symmetry \cite{gorshkov_two-orbital_2010, cazalilla_ultracold_2014}.
In contrast to the well-studied SU(2) FHM, much less is known about the extended SU($N>2$)-symmetric case. Calculations are mostly limited to $T=0$, low dimensions or approximated correlations, but point to very rich phase diagrams including novel magnetic and spin liquid phases
\cite{honerkamp_ultracold_2004,Cherng2007,hermele_mott_2009,gorshkov_two-orbital_2010,Rapp2011,Inaba2013,Song2013,sotnikov_magnetic_2014,Sotnikov2015}. Moreover, accurate predictions of thermodynamic quantities are even harder to obtain than for the SU(2) case,
as most numerical algorithms fail for larger $N$ due the unfavourable scaling of the Hilbert space.

Fermionic ${}^{173}$Yb has nuclear spin $I=5/2$, but no electronic angular momentum in the ground state ($J=0$). Therefore, nuclear and electronic angular momenta are decoupled, making the atomic interactions independent of the nuclear spin state and the system  SU($N$)-symmetric, with $N\leq 2I+1=6$ being the number of populated nuclear spin states \cite{scazza_observation_2014}. One important manifestation of large $N$ is an enhanced Pomeranchuk effect, leading to a suppression of particle-hole excitations in the lattice \cite{hazzard_high-temperature_2012}. Owing to this, first evidence of an incompressible phase in an SU(6) $^{173}$Yb gas has recently been reported, at an entropy level that does not support a Mott insulator in lower spin gases \cite{taie_su6_2012}.

The SU($N$)-symmetric FHM for arbitrary spin multiplicity $N$ can be written as \cite{cazalilla_ultracold_2014}
\[
\hat{H}=-t{\displaystyle \sum_{\langle i,j\rangle,\sigma}(\hat{c}_{i\sigma}^{\dagger}\hat{c}_{j\sigma}^{}+{\rm H.c.})}+\frac{U}{2}{\displaystyle \sum_{i,\sigma\neq\sigma'}\hat{n}_{i\sigma}\hat{n}_{i\sigma'}}+{\displaystyle \sum_{i,\sigma}V_{i}\hat{n}_{i\sigma}.}
\]
Here, $\langle i,j \rangle$ denote neighbouring lattice site indices, $t$ is the tunnelling matrix element between them, $U$ is the on-site interaction and $V_i$ is a position-dependent energy offset which accounts for the confining potential. We denote with $t^*=12\,t$ the kinetic energy associated with the bandwidth of the 3D lattice. The operator $ \hat{c}_{i\sigma}$ annihilates a fermion at site $i$ with spin index $\sigma=1, ... , N$ and $\hat{n}_{i\sigma}=\hat{c}^\dagger_{i\sigma}\hat{c}^{}_{i\sigma}$ are the respective number operators. 

In this work, we measure and analyse the equation of state (EoS) of an SU($N$) Fermi gas in a cubic optical lattice, for temperatures above the magnetic ordering temperature. The EoS is a thermodynamic relation that contains all the macroscopic properties of the system and, due to its generality, is particularly well suited to benchmark numerical simulations \cite{nascimbene_exploring_2010,VanHoucke2012}.
We take advantage of the confining potential and the local density approximation (LDA) to map the trapped heterogeneous gas with fixed number of particles to a locally homogeneous gas in the grand canonical ensemble with $\mu_i = \mu_0 - V_i$, where $\mu_0$ is the chemical potential at the centre of the trap.
The validity of the LDA for fermions trapped in optical lattices above the Néel temperature was previously verified by numerical calculations \cite{helmes_mott_2008}.
We determine the EoS for the density $n(\mu,T,N,U,t^*)$ over a wide range of parameters. In particular, we focus on the highest spin multiplicity of our system $N=6$ and on the case $N=3$, which was the subject of several theoretical studies \cite{gorelik_mott_2009, Rapp2011, Song2013, Inaba2013, sotnikov_magnetic_2014, Sotnikov2015}. By deriving the local compressibility directly from the measured EoS we are able to detect the emergence of the incompressible Mott phase.


In our experiment, we start by preparing a degenerate Fermi gas of $^{173}$Yb with initially $N=6$ equally populated spin components via evaporative cooling in a crossed optical dipole trap (see Methods for details). We then set $N$ by removing individual spin components. The Fermi gas is then loaded into the lowest energy band of a 3D optical lattice with cubic symmetry and lattice spacing $d=\lambda /2$ operating at a wavelength of $\lambda = 759$\,nm. We vary the lattice depth between $3\,E_{\rm r}$ and $15\,E_{\rm r}$, with $E_{\rm r}=h^2/2m\lambda^2$ being the recoil energy, a range for which the tight-binding approximation is valid. Adjusting the lattice depth allows us to tune the system from $U/t^*=0.128^{+0.004}_{-0.008}$ to $U/t^*=11.0^{+1.1}_{-1.0}$, spanning a range of two orders of magnitude.


\begin{figure}[t]
	\includegraphics[width=0.48\textwidth]{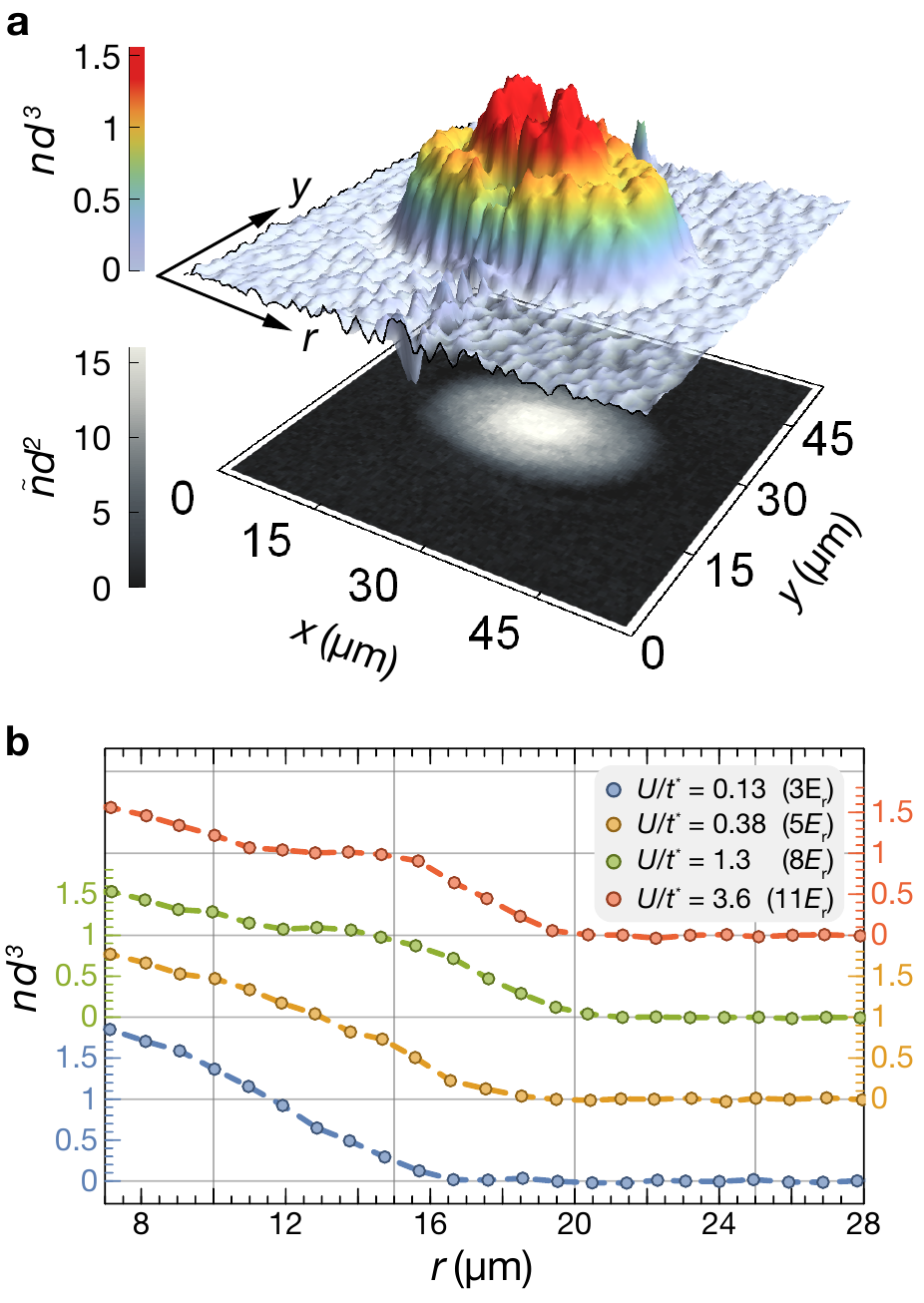} 
	\caption{\label{fig:atoms_waterfall}\textbf{Density profile of an SU($6$) Fermi gas in an optical lattice with harmonic confinement}. \textbf{a}, measured density distribution integrated along the line of sight (bottom) and local trap density obtained by performing the inverse Abel transform, displaying a plateau $n(r,y)=1$  (top) for interaction strength $U/t^*=6.4$ ($V=13E_{\rm r}$). Three experimental realizations are averaged for the displayed data.
		\textbf{b}, local atomic density as a function of the distance to the trap center for different interaction parameters $U/t^*$.
	}
	\vspace{\capVert}
\end{figure}

In order to measure the local atomic density in the trap, we probe the cloud by performing \textit{in situ} spin-independent absorption imaging along the $z$-axis of the lattice. The spatial resolution is approximately $1.2\,\mu \rm{m}$ $ \approx 3.2\,d$. 
Due to the high optical density of the trapped cloud, saturated absorption imaging at high light intensity is used (see Methods and ref. \cite{reinaudi_strong_2007} for details). 
The resulting integrated two-dimensional density distribution $\tilde{n}(x,y)$ is shown in Fig.  \ref{fig:atoms_waterfall}a. 
Exploiting the carefully characterized geometry of the trap configuration, we determine the local 3D density $n(r,y)$ by performing an inverse Abel transform with $y$ as the symmetry axis. \cite{dribinski_reconstruction_2002}. The transformation is stable for pixels with radial distance $r\ge 7$\,$\mu m$ from the symmetry axis.
In Fig. \ref{fig:atoms_waterfall}b, the azimuthally averaged density is shown as a function of the distance to the trap centre for different values of the lattice depth. 
For the averaging, data within $\pm4.8\,\mu$m along the $y$-axis was used. 

Two regimes can be distinguished: For low lattice depths, where the interaction energy is smaller than the kinetic energy ($U < t^*$), the density decreases smoothly from the centre to the edge of the trap, indicating that the system is compressible everywhere. 
For high lattice depths ($U> t^*$), we observe the formation of a plateau of constant density, as expected for a Mott insulating state\ \cite{duarte_compressibility_2015}.
In this region, the atomic density computed using an independent calibration is compatible with the expected value of one fermion per lattice site (see Methods). Due to the higher precision of this measurement, the density of an SU(6) insulator in this region is used as the density reference in this paper.
	
\begin{figure}[!p]
	\centering
	\includegraphics[width=0.48\textwidth]{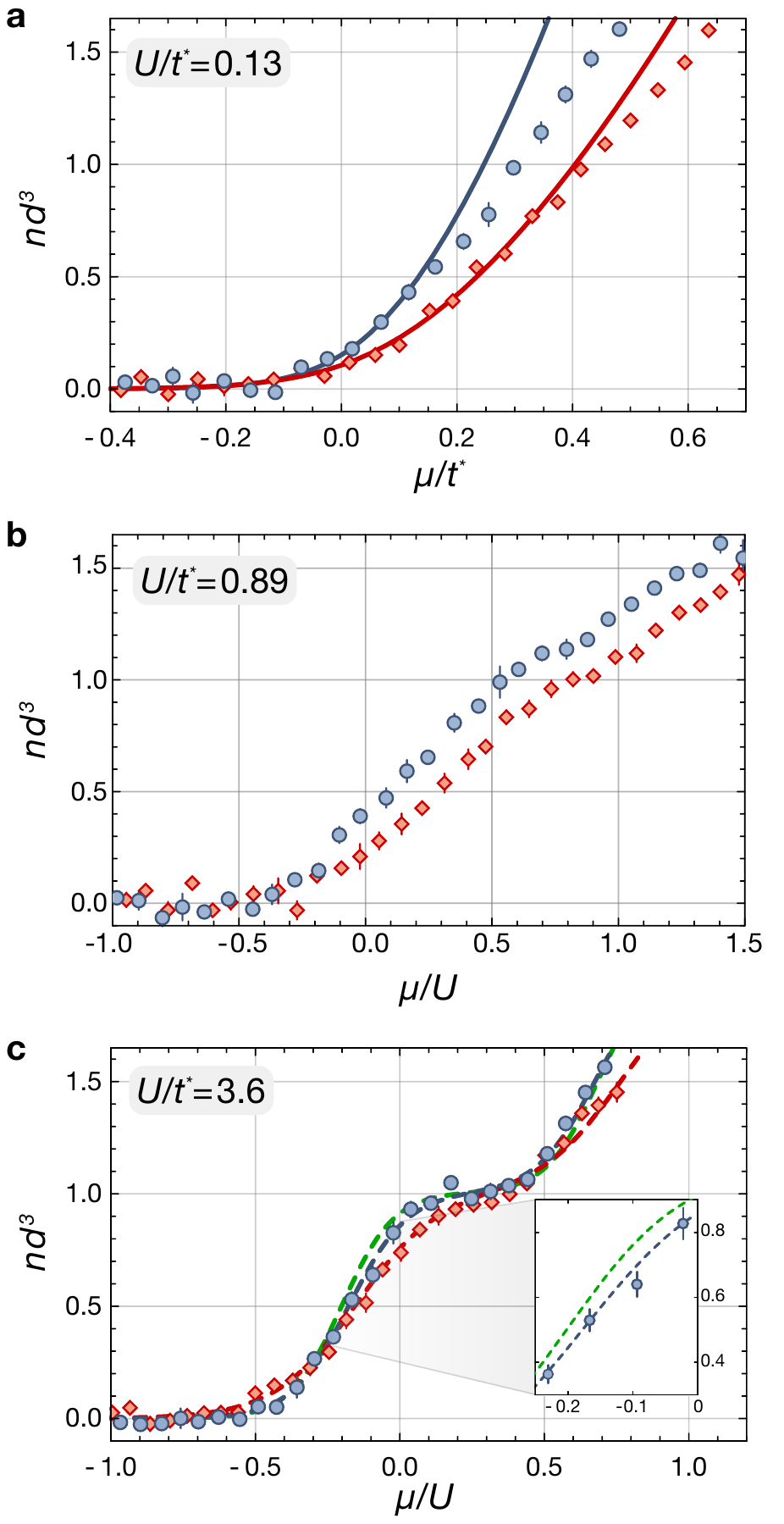} 
	
	\caption{\label{fig:NLattice}\label{fig:EOS}\textbf{EoS of the SU($6$)- (blue circles) and the  SU($3$)- (red diamonds) spin symmetric Fermi gas in a lattice.} Density as a function of the chemical potential for various interaction strengths: \textbf{a}, $U/t^*=0.128$ \textbf{b}, $U/t^*=0.89$ and \textbf{c}, $U/t^*=3.6$. Solid lines are fits to the non-interacting Fermi gas EoS, with points $nd^3<0.5$ included in the fit.
		Dashed lines are the fits to the low tunnelling model for SU($6$) (blue) and SU($3$) (red). An SU($2$) model fit to the SU($6$) data is shown for comparison (green), with the temperature $T$ and the central chemical potential $\mu_0$ as fit parameters. Error bars are the s.e.m. of the binned data.}
	\hspace{-9ex}
	
\end{figure}

To determine thermodynamic quantities, the EoS is obtained by relating the 3D atom density distribution and the chemical potential: $n[\mu = \mu_0-V(r,y)]$. This is done for different values of interaction strength $U/t^*$ and number of spin components $N=3$, 6. In Fig. \ref{fig:EOS}, examples for small ($U/t^*=0.128$), intermediate ($U/t^*=0.89$) and large ($U/t^*=3.6$) interaction strengths are given.


In the presence of a weak lattice ($U\ll t^*$), the system is in a normal metallic state, well described by the SU($N$) Fermi-liquid theory \cite{yip_theory_2014} with enhanced mass and interactions due to the presence of the lattice. In particular, for low densities $nd^3 \ll 1$, interactions can be neglected and the EoS is well approximated by that of a non-interacting Fermi gas, as shown in Fig. \ref{fig:NLattice}a. We use this observation to determine $\mu_0$ of the gases with $U/t^* \leq 0.89$ (see Methods for details). For high densities $nd^3 \gtrsim 1$ the approximation fails and an SU($N$) theory that considers interactions in a weak lattice would be required.

For deep lattice potentials, tunnelling is highly suppressed ($t\ll k_{\rm{B}} T$) and interactions are strong ($U \gg t^*$). In this regime, the lattice sites can approximately be regarded as independent. We construct a low-tunnelling model using a high-temperature series expansion (HTSE) up to  $\mathcal{O} (t/k_{\rm{B}} T)^2$ \cite{Oitmaa2010,hazzard_high-temperature_2012}  (see Methods for details). This model fits the measured EoS data well, with $T=0.13(1)\,U/k_{\rm{B}}$ and $T=0.18(1)\,U/k_{\rm{B}}$ for SU($6$) and SU($3$) respectively (see Fig. \ref{fig:NLattice}c). The lower temperature in the SU(6) case is due to both the lower initial temperature and the stronger Pomeranchuk effect. As the HTSE fit sensitivity to the trap confinement is higher than the precision of the independent trap calibration, we allow a variation of the confinement parameter in the model within the bounds given by the calibration.


\begin{figure*}[t]
	\includegraphics[width=0.9\textwidth]{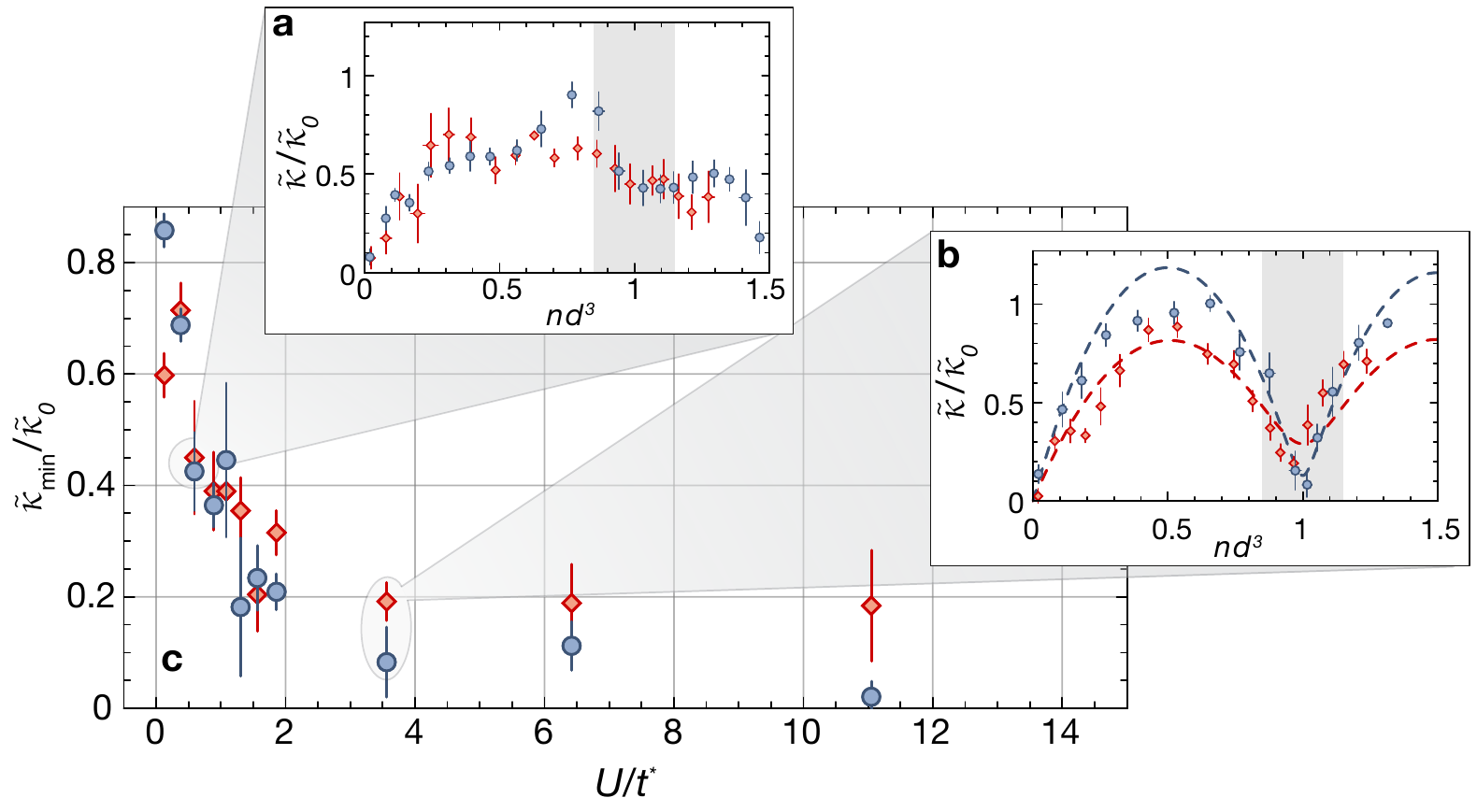}
	\caption{\label{fig:MottTransition}
		\textbf{Compressibility of SU($6$)- (blue circles) and SU($3$)- (red diamonds) spin-symmetric Fermi gases in a lattice.}
		\textbf{a}, compressibility as a function of density in the intermediate regime $U/t^*=0.59$.
		\label{fig:compHighLattice} \label{fig:com} \textbf{b}, compressibility as a function of density in the strongly interacting regime $U/t^*=3.6$. The dashed lines are obtained by deriving the fitted low-tunnelling model plotted in Fig. \ref{fig:NLattice}\textbf{c}.
		\textbf{c}, minimal compressibility $\tilde{\kappa}_{\rm{min}}/\tilde{\kappa}_0$ as a function of interaction strength, where $\tilde{\kappa}_0$ is the compressibility of a non-interacting SU($6$) Fermi gas at $T=0$ and $n=1/d^3$. The minimum of the compressibility  $\tilde{\kappa}(n)$ in an interval around unit filling (shaded area in insets) is used as $\tilde{\kappa}_{\rm{min}}$.
		Error bars denote the confidence intervals of the linear regression used to obtain $\tilde{\kappa}$.}
	\vspace{\capVert}
\end{figure*}

The role of the SU($N$) symmetry in the EoS is simple in the limit of negligible interactions: the density scales linearly with $N$ for fixed chemical potential. This dependence is well reproduced by the data as shown in Fig. \ref{fig:NLattice}a. For high lattice depths, and therefore larger $U/t^*$, the effect is two-fold: Firstly, the Pomeranchuk effect leads to lower temperatures in systems with higher $N$ \cite{hazzard_high-temperature_2012,taie_su6_2012}. Secondly, for a given finite temperature, the EoS in the strongly interacting regime is $N$-dependent in a non-trivial way due to the different quantum statistics \cite{sotnikov_magnetic_2014}. To illustrate this, we compare the low-tunnelling model for the SU($2$) case and the EoS for SU($6$) in Fig. \ref{fig:NLattice}c. As shown in the inset, a measurable discrepancy is present due to the different $N$-dependent quantum statistics, which cannot be attributed to different temperature $T$ or central chemical potential $\mu_0$.

In the regime of intermediate interaction strength, comparable to the kinetic energy ($U\approx t^*$), the system is in a strongly correlated many-body state. No exact determination of the EoS is available yet to compare to our data, displayed in Fig. \ref{fig:NLattice}b.
Nevertheless, having model-free access to the EoS allows to directly determine the local  compressibility $\tilde{\kappa}=n^2 \kappa=\partial n/\partial \mu |_T$.
This can be used to probe the emergence of the incompressible phase and to study the crossover between metal and Mott insulator, as shown in Fig. \ref{fig:MottTransition}. For the strongly interacting case we distinguish a metallic outer layer $0<nd^3<1$ and a metallic core $1<nd^3$, separated by a Mott shell with $nd^3=1$. In this regime, the low-tunnelling model compares well with our results, with very small compressibilities in the Mott regions. 

To characterize the onset of the Mott phase, we determine the minimum of compressibility $\tilde{\kappa}_{\rm{min}}$ around unit filling, in a range $nd^3 \in [0.85, 1.15]$, as a function of interaction parameter (Fig. \ref{fig:MottTransition}). We observe a suppression of $\tilde{\kappa}_{\rm{min}}$ by roughly one order of magnitude when increasing the interaction strength. For large $U/t^*$ the compressibility saturates at a minimum value and the Mott shell is formed. The behaviour is consistent with numerical calculations \cite{gorelik_mott_2009}.

\begin{figure}[t]
	\includegraphics[width=0.48\textwidth]{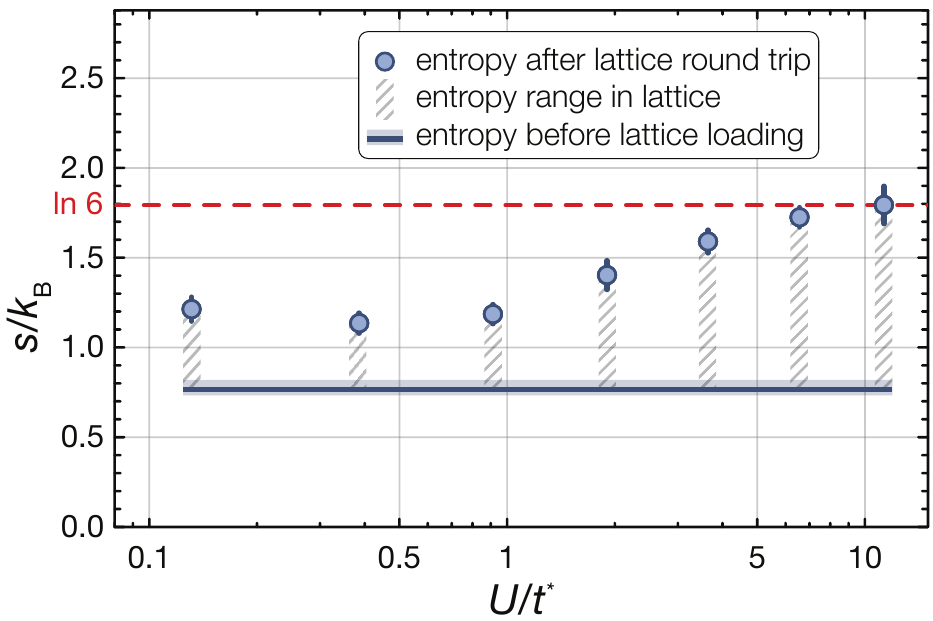}	
	\caption{\label{fig:entropy}\textbf{Entropy per particle $s$ of the SU($6$)-symmetric gas before and after the lattice round-trip sequence} The blue line represents the entropy measured in the bulk before transfer into the lattice. Blue points are measurements after round-trip loading into the lattice and back to the bulk. The entropy of the gas in the lattice is constrained to the region between these two measurements. The entropy of a system with our parameters and fully random spin orientations is close to  $s/k_\mathrm{B}=$\,\,{ln 6} (red line). Error bars and blue shaded region are the s.e.m. of the temperature measurements.}
	\vspace{\capVert}
\end{figure}

To further characterize the state in the lattice, we estimate the entropy per particle $s$ of the SU($6$) gas. We determine the entropy before and after a round trip, consisting of loading the atoms into the lattice and back into the dipole trap, by measuring the degeneracy parameter $T/T_{\rm F}$ in the dipole trap, where $T_{\rm F}$ is the Fermi temperature. 
This gives a lower and an upper estimate 
for $s$ in the lattice (see Fig. \ref{fig:entropy}). For low lattice depths, we observe a constant entropy rise associated with the transfer from the bulk into the lattice and back.  Above $U \approx t^*$, with the appearance of the incompressible phase, we find an additional entropy increase. This increase also persists for very low ramp speeds, which could indicate diminished adiabaticity due to the insulating phase crossover hindering mass flow. However, we observe that also for these high lattice depths, the entropy is at or below the maximum spin entropy even
after a full round trip that includes the reversed second ramp. This indicates that the orientations of the spins in the lattice are likely not fully random, allowing for the presence of partial spin correlations in the system.
	
In the regime of validity of the low-tunnelling model, lower entropy corresponds to lower temperature and Mott shells with sharper edges. Although fitting the model provides a good match to the measured SU(6)-EoS data as shown in Fig. \ref{fig:NLattice}c, the obtained temperature is significantly higher than that expected for the very low entropy measured in the round-trip experiment.
In contrast, in the SU(3) case the measured round-trip entropy is higher than the saturated spin entropy and the temperature obtained from the \textit{in situ} HTSE fit is in the expected range.
	
An interpretation for the SU(6) case could be that the suppression of mass flow close to the Mott crossover prevents the formation of sharp Mott shell edges, yielding higher fitted temperatures.
Nevertheless, the measured low entropy after the round-trip hints that this non-adiabatic process would have to be mostly reversible, except for the observed increase seen in Fig. \ref{fig:entropy}. Understanding a possible edge-softening contribution of entropy stored in that metallic region would require a quantitative solution of the SU(6) 3D theory at low filling and including spin correlations, a  challenge for current numerical methods.


In conclusion, we use \textit{in situ} probing to measure the equation of state of an SU($N$)-symmetric Fermi gas in a 3D lattice. We
obtain a very low compressibility in the Mott insulating phase and measure an entropy below that of uncorrelated spins.
These findings make the system a promising starting point towards novel magnetically ordered many-body states in highly spin-symmetric systems \cite{honerkamp_ultracold_2004, Cherng2007, yip_theory_2014, cazalilla_ultracold_2014,sotnikov_magnetic_2014,Sotnikov2015}.
An incompressible phase with filling $nd^3=1$ is expected to faithfully realize the SU($N$)-symmetric Heisenberg model, paving the way towards studying unexplored systems with reduced dimensions such as chains \cite{Manmana2011,Messio2012,Capponi2015} and new magnetic phases in square lattices \cite{honerkamp_ultracold_2004}.

\section{Methods}
\subsection{Preparation of SU(\textit{N}) Fermi gases}
Approximately 3.5 million ground-state atoms of $^{173}$Yb, with equal  populations in the nuclear spin states, are loaded into a crossed optical dipole trap operating at a wavelength of $\lambda = 1064$\,nm. Forced evaporative cooling produces a degenerate Fermi gas of 5000 atoms per spin state at temperature $T=0.07(1)\,T_{\mathrm{F}}$, where $T_\mathrm{F}$ is the Fermi temperature. The gas is very weakly interacting since $k_{\rm F}a_s \lesssim 0.07$, with $a_s=199.4\,a_0$ being the \textit{s}-wave scattering length and $a_0$ the Bohr radius. We perform state preparation by driving the $^{1}\mathrm{S}_0 \to {}^{3}\mathrm{P}_1$ optical transition to remove unwanted nuclear spin states from the trap, in the presence of a homogeneous magnetic field that lifts the spin state degeneracy. Using this technique we generate an SU($3$) Fermi gas with a temperature of $0.15(1)\,T_{\rm{F}}$, with a residual fraction of unwanted spin components below 5$\%$ per component.

\subsection{Loading into the optical lattice}
At the end of evaporation, the atoms are transferred into a cubic optical lattice in two steps. The sample is first loaded into a shallow lattice with depth $V=3\,E_{\rm r}$ in $150\,$ms, avoiding band excitations. We then ramp the lattice depth to the final value, between $V=3\,E_{\rm r}$ and $V=15\,E_{\rm r}$, in $150\,$ms. The atoms are trapped in the combined harmonic confinement produced by the lattice beams and the crossed optical dipole trap. The confinement frequencies vary between $\omega_{x,y,z}= 2 \pi \times  (31,42,183)\,$Hz at $3\,E_{\rm r}$, $\omega_{x,y,z}= 2 \pi \times  (21,33,183)\,$Hz at $7\,E_{\rm r}$ and are approximately constant for $V> 7\,E_{\rm r}$.
We verified the validity of the harmonic approximation for our experimental trap configuration, by taking into account the combined Gaussian beam profiles. The variation of $U/t^*$ in the region occupied by the atoms is estimated to be below 8\%.

Dilute samples are loaded into the lattice in order to minimize losses and heating from three-body recombination, which is not suppressed by Pauli blocking for $N>2$ (see Supplementary Information for more details).

\subsection{High-intensity imaging}
The atom cloud has a typical optical density around 2 in the trap centre. In order to perform \textit{in situ} absorption imaging we saturate the imaging transition $^{1}\mathrm{S}_0 \to {}^{1}\mathrm{P}_1$ with light intensity $I=15\,I_{\rm{sat}}$, where $I_{\rm{sat}} = 60\,$mW/cm$^2$. Details on the calibration procedure are in the Supplementary Information. The optical resolution is about $1.2\,\mu$m, determined as described in ref. \cite{Hung2011}.
The imaging pulse has a duration of $5\,\mu$s, sufficiently short to avoid atoms to escape from the focal plane and to minimize the Doppler shift due to photon scattering. Imaging is performed in the absence of magnetic fields in order to have spin-independent detection. Nevertheless, a 6\% difference between the densities of the Mott plateaus is found comparing the SU($6$) and SU($3$) gases, likely caused by the different line strengths within the hyperfine substructure of the imaging transition.

\subsection{Analytical models}
In the absence of interactions, the EoS of a $N$-component harmonically confined Fermi gas is
\[
n(\mu,T)=N \left(\frac{m k_{\mathrm{B}}T}{2\pi\hbar^2}\right)^{3/2}  {\rm{Li}}_{3/2} \left(-e^{\mu/k_{\rm B}T}\right),
\]
where $\rm{Li}_{3/2}$ denotes the poly-logarithm of order 3/2. This expression is also valid in the presence of a weak lattice potential, provided that the effective mass associated with the dispersion of the lowest band is used.

In the strongly interacting regime, the low-tunnelling model is obtained using a high-temperature series expansion (HTSE) of the grand canonical potential up to second order in $t/k_{\rm{B}}T$ \cite{Oitmaa2010,hazzard_high-temperature_2012,taie_su6_2012}:
\begin{eqnarray*}
\Omega & = & \Omega_{0}+\Delta\Omega
\end{eqnarray*}
\begin{eqnarray*}
\Omega_{0}(\mu,T) & = & -(1/\beta)\ln Z_{0}(T,\mu)\\
Z_{0}(\mu,T) & = & \sum_{n=0}^{N}{\binom{N}{n}}e^{-\beta\left(\frac U 2 n(n-1)-\mu n\right)}
\end{eqnarray*}
\begin{gather*}
-\beta\Delta\Omega=\left(\frac{\beta t}{Z_{0}}\right)^{2}zN\left[\frac{1}{2}\sum_{n_{1}=1}^{N}{\binom{N-1}{n_{1}-1}}^{2}x^{2n_{1}-1}y^{(n_{1}-1)^{2}}\right.\\
-\frac{1}{\beta U}\sum_{n_{1}\neq n_{2}}^{N}{\binom{N-1}{n_{1}-1}}{\binom{N-1}{n_{2}-1}}\\
\left.\frac{x^{n_{1}+n_{2}-1}y^{\frac{1}{2}n_{1}(n_{1}-1)+\frac{1}{2}(n_{2}-1)(n_{2}-2)}}{n_{1}-n_{2}}\right]
\end{gather*}
with $z$ being the number of next neighbours in the lattice, $x=e^{\beta\mu}$, $y=e^{-\beta U}$ and $\Omega_0(\mu,T)$ the grand potential and $Z_0$ the partition function in the atomic limit.

\subsection{Fitting and calibration}
The effective optical scattering cross section $\sigma$ of the atoms is calibrated using the known EoS of the polarized Fermi gas in the dipole trap with known trap frequencies.
We obtain $\sigma$/$\sigma_0=0.222 \pm 0.034$ with $\sigma_0=\frac{3}{2\pi}\lambda^2$.

The HTSE model fit to the measured EoS, effectively fitting the amplitude of the plateau, is more sensitive to the optical cross section and yields $\sigma$/$\sigma_0=0.233 \pm 0.003$, which is used in this work. The plateau density for the $N=6$ data shown is therefore fixed to $1/d^3$, whereas an independent fit for the $N=3$ case yields $nd^3=0.94 \pm 0.02$, possibly due to residual spin dependence of the imaging technique.

\vspace{2mm}

\section{Additional information}
\noindent
Correspondence and requests for materials should be addressed to S.F.

\section{Acknowledgements}
We acknowledge fruitful discussions with Christophe Salomon and Tarik Yefsah. This work was supported by the ERC through the synergy grant UQUAM and by the European Union's Horizon 2020 funding (D.R.F.).


\noindent






\begin{thebibliography}{10}
\expandafter\ifx\csname url\endcsname\relax
  \def\url#1{\texttt{#1}}\fi
\expandafter\ifx\csname urlprefix\endcsname\relax\def\urlprefix{URL }\fi
\providecommand{\bibinfo}[2]{#2}
\providecommand{\eprint}[2][]{\url{#2}}

\bibitem{Imada1998}
\bibinfo{author}{Imada, M.}, \bibinfo{author}{Fujimori, A.} \&
  \bibinfo{author}{Tokura, Y.}
\newblock \bibinfo{title}{Metal-insulator transitions}.
\newblock \emph{\bibinfo{journal}{Rev. Mod. Phys.}}
  \textbf{\bibinfo{volume}{70}}, \bibinfo{pages}{1039--1263}
  (\bibinfo{year}{1998}).

\bibitem{Tokura2000}
\bibinfo{author}{Tokura, Y.} \& \bibinfo{author}{Nagaosa, N.}
\newblock \bibinfo{title}{Orbital physics in transition-metal oxides}.
\newblock \emph{\bibinfo{journal}{Science}} \textbf{\bibinfo{volume}{288}},
  \bibinfo{pages}{462--468} (\bibinfo{year}{2000}).

\bibitem{cazalilla_ultracold_2014}
\bibinfo{author}{Cazalilla, M.~A.} \& \bibinfo{author}{Rey, A.~M.}
\newblock \bibinfo{title}{Ultracold {Fermi} gases with emergent {SU}({$N$})
  symmetry}.
\newblock \emph{\bibinfo{journal}{Rep. Prog. Phys.}}
  \textbf{\bibinfo{volume}{77}}, \bibinfo{pages}{124401}
  (\bibinfo{year}{2014}).

\bibitem{Anderson1987}
\bibinfo{author}{Anderson, P.~W.}
\newblock \bibinfo{title}{The resonating valence bond state in La2CuO4 and
  superconductivity}.
\newblock \emph{\bibinfo{journal}{Science}} \textbf{\bibinfo{volume}{235}},
  \bibinfo{pages}{1196--1198} (\bibinfo{year}{1987}).

\bibitem{Esslinger2010}
\bibinfo{author}{Esslinger, T.}
\newblock \bibinfo{title}{Fermi-Hubbard physics with atoms in an optical
  lattice}.
\newblock \emph{\bibinfo{journal}{Annu. Rev. Condens. Matter Phys.}}
  \textbf{\bibinfo{volume}{1}}, \bibinfo{pages}{129--152}
  (\bibinfo{year}{2010}).

\bibitem{gorshkov_two-orbital_2010}
\bibinfo{author}{Gorshkov, A.~V.} \emph{et~al.}
\newblock \bibinfo{title}{Two-orbital {SU}({$N$}) magnetism with ultracold
  alkaline-earth atoms}.
\newblock \emph{\bibinfo{journal}{Nat. Phys.}} \textbf{\bibinfo{volume}{6}},
  \bibinfo{pages}{289--295} (\bibinfo{year}{2010}).

\bibitem{honerkamp_ultracold_2004}
\bibinfo{author}{Honerkamp, C.} \& \bibinfo{author}{Hofstetter, W.}
\newblock \bibinfo{title}{Ultracold fermions and the $\mathrm{SU}(N)$ Hubbard
  model}.
\newblock \emph{\bibinfo{journal}{Phys. Rev. Lett.}}
  \textbf{\bibinfo{volume}{92}}, \bibinfo{pages}{170403}
  (\bibinfo{year}{2004}).

\bibitem{Cherng2007}
\bibinfo{author}{Cherng, R.~W.}, \bibinfo{author}{Refael, G.} \&
  \bibinfo{author}{Demler, E.}
\newblock \bibinfo{title}{Superfluidity and magnetism in multicomponent
  ultracold fermions}.
\newblock \emph{\bibinfo{journal}{Phys. Rev. Lett.}}
  \textbf{\bibinfo{volume}{99}}, \bibinfo{pages}{130406}
  (\bibinfo{year}{2007}).

\bibitem{hermele_mott_2009}
\bibinfo{author}{Hermele, M.}, \bibinfo{author}{Gurarie, V.} \&
  \bibinfo{author}{Rey, A.~M.}
\newblock \bibinfo{title}{Mott insulators of ultracold fermionic
  alkaline earth atoms: underconstrained magnetism and chiral
  spin liquid}.
\newblock \emph{\bibinfo{journal}{Phys. Rev. Lett.}}
  \textbf{\bibinfo{volume}{103}}, \bibinfo{pages}{135301}
  (\bibinfo{year}{2009}).

\bibitem{Rapp2011}
\bibinfo{author}{Rapp, A.} \& \bibinfo{author}{Rosch, A.}
\newblock \bibinfo{title}{Ground-state phase diagram of the repulsive SU(3)
  Hubbard model in the Gutzwiller approximation}.
\newblock \emph{\bibinfo{journal}{Phys. Rev. A}} \textbf{\bibinfo{volume}{83}},
  \bibinfo{pages}{053605} (\bibinfo{year}{2011}).

\bibitem{Inaba2013}
\bibinfo{author}{Inaba, K.} \& \bibinfo{author}{Suga, S.-I.}
\newblock \bibinfo{title}{Superfluid, staggered state, and Mott insulator of
  repulsively interacting three-component fermionic atoms in optical lattices}.
\newblock \emph{\bibinfo{journal}{Mod. Phys. Lett. B}}
  \textbf{\bibinfo{volume}{27}}, \bibinfo{pages}{1330008}
  (\bibinfo{year}{2013}).

\bibitem{Song2013}
\bibinfo{author}{Song, H.} \& \bibinfo{author}{Hermele, M.}
\newblock \bibinfo{title}{Mott insulators of ultracold fermionic alkaline earth
  atoms in three dimensions}.
\newblock \emph{\bibinfo{journal}{Phys. Rev. B}} \textbf{\bibinfo{volume}{87}},
  \bibinfo{pages}{1--9} (\bibinfo{year}{2013}).

\bibitem{sotnikov_magnetic_2014}
\bibinfo{author}{Sotnikov, A.} \& \bibinfo{author}{Hofstetter, W.}
\newblock \bibinfo{title}{Magnetic ordering of three-component ultracold
  fermionic mixtures in optical lattices}.
\newblock \emph{\bibinfo{journal}{Phys. Rev. A}} \textbf{\bibinfo{volume}{89}},
  \bibinfo{pages}{063601} (\bibinfo{year}{2014}).

\bibitem{Sotnikov2015}
\bibinfo{author}{Sotnikov, A.}
\newblock \bibinfo{title}{{Critical entropies and magnetic-phase-diagram
  analysis of ultracold three-component fermionic mixtures in optical
  lattices}}.
\newblock \emph{\bibinfo{journal}{Phys. Rev. A}} \textbf{\bibinfo{volume}{92}},
  \bibinfo{pages}{023633} (\bibinfo{year}{2015}).

\bibitem{scazza_observation_2014}
\bibinfo{author}{Scazza, F.} \emph{et~al.}
\newblock \bibinfo{title}{Observation of two-orbital spin-exchange interactions
  with ultracold {SU}({$N$})-symmetric fermions}.
\newblock \emph{\bibinfo{journal}{Nat. Phys.}} \textbf{\bibinfo{volume}{10}},
  \bibinfo{pages}{779--784} (\bibinfo{year}{2014}).

\bibitem{hazzard_high-temperature_2012}
\bibinfo{author}{Hazzard, K. R.~A.}, \bibinfo{author}{Gurarie, V.},
  \bibinfo{author}{Hermele, M.} \& \bibinfo{author}{Rey, A.~M.}
\newblock \bibinfo{title}{High-temperature properties of fermionic
  alkaline-earth-metal atoms in optical lattices}.
\newblock \emph{\bibinfo{journal}{Phys. Rev. A}} \textbf{\bibinfo{volume}{85}},
  \bibinfo{pages}{041604} (\bibinfo{year}{2012}).

\bibitem{taie_su6_2012}
\bibinfo{author}{Taie, S.}, \bibinfo{author}{Yamazaki, R.},
  \bibinfo{author}{Sugawa, S.} \& \bibinfo{author}{Takahashi, Y.}
\newblock \bibinfo{title}{An {SU}(6) {Mott} insulator of an atomic {Fermi} gas
  realized by large-spin {Pomeranchuk} cooling}.
\newblock \emph{\bibinfo{journal}{Nat. Phys.}} \textbf{\bibinfo{volume}{8}},
  \bibinfo{pages}{825--830} (\bibinfo{year}{2012}).

\bibitem{nascimbene_exploring_2010}
\bibinfo{author}{Nascimbène, S.}, \bibinfo{author}{Navon, N.},
  \bibinfo{author}{Jiang, K.~J.}, \bibinfo{author}{Chevy, F.} \&
  \bibinfo{author}{Salomon, C.}
\newblock \bibinfo{title}{Exploring the thermodynamics of a universal {Fermi}
  gas}.
\newblock \emph{\bibinfo{journal}{Nature}} \textbf{\bibinfo{volume}{463}},
  \bibinfo{pages}{1057--1060} (\bibinfo{year}{2010}).

\bibitem{VanHoucke2012}
\bibinfo{author}{{Van Houcke}, K.} \emph{et~al.}
\newblock \bibinfo{title}{{Feynman diagrams versus Fermi-gas Feynman
  emulator}}.
\newblock \emph{\bibinfo{journal}{Nat. Phys.}} \textbf{\bibinfo{volume}{8}},
  \bibinfo{pages}{366--370} (\bibinfo{year}{2012}).

\bibitem{helmes_mott_2008}
\bibinfo{author}{Helmes, R.~W.}, \bibinfo{author}{Costi, T.~A.} \&
  \bibinfo{author}{Rosch, A.}
\newblock \bibinfo{title}{Mott transition of fermionic atoms in a three-dimensional optical trap}.
\newblock \emph{\bibinfo{journal}{Phys. Rev. Lett.}}
  \textbf{\bibinfo{volume}{100}}, \bibinfo{pages}{056403}
  (\bibinfo{year}{2008}).

\bibitem{gorelik_mott_2009}
\bibinfo{author}{Gorelik, E.~V.} \& \bibinfo{author}{Blümer, N.}
\newblock \bibinfo{title}{Mott transitions in ternary flavor mixtures of
  ultracold fermions on optical lattices}.
\newblock \emph{\bibinfo{journal}{Phys. Rev. A}} \textbf{\bibinfo{volume}{80}},
  \bibinfo{pages}{051602} (\bibinfo{year}{2009}).

\bibitem{reinaudi_strong_2007}
\bibinfo{author}{Reinaudi, G.}, \bibinfo{author}{Lahaye, T.},
  \bibinfo{author}{Wang, Z.} \& \bibinfo{author}{Guéry-Odelin, D.}
\newblock \bibinfo{title}{Strong saturation absorption imaging of dense clouds
  of ultracold atoms}.
\newblock \emph{\bibinfo{journal}{Opt. Lett.}} \textbf{\bibinfo{volume}{32}},
  \bibinfo{pages}{3143} (\bibinfo{year}{2007}).

\bibitem{dribinski_reconstruction_2002}
\bibinfo{author}{Dribinski, V.}, \bibinfo{author}{Ossadtchi, A.},
  \bibinfo{author}{Mandelshtam, V.~A.} \& \bibinfo{author}{Reisler, H.}
\newblock \bibinfo{title}{Reconstruction of {Abel}-transformable images: {The}
  {Gaussian} basis-set expansion {Abel} transform method}.
\newblock \emph{\bibinfo{journal}{Rev. Sci. Instrum.}}
  \textbf{\bibinfo{volume}{73}}, \bibinfo{pages}{2634--2642}
  (\bibinfo{year}{2002}).

\bibitem{duarte_compressibility_2015}
\bibinfo{author}{Duarte, P.~M.} \emph{et~al.}
\newblock \bibinfo{title}{Compressibility of a fermionic Mott insulator of
  ultracold atoms}.
\newblock \emph{\bibinfo{journal}{Phys. Rev. Lett.}}
  \textbf{\bibinfo{volume}{114}}, \bibinfo{pages}{070403}
  (\bibinfo{year}{2015}).

\bibitem{yip_theory_2014}
\bibinfo{author}{Yip, S.-K.}, \bibinfo{author}{Huang, B.-L.} \&
  \bibinfo{author}{Kao, J.-S.}
\newblock \bibinfo{title}{Theory of $\mathrm{SU}(N)$ Fermi liquids}.
\newblock \emph{\bibinfo{journal}{Phys. Rev. A}} \textbf{\bibinfo{volume}{89}},
  \bibinfo{pages}{043610} (\bibinfo{year}{2014}).

\bibitem{Oitmaa2010}
\bibinfo{author}{Oitmaa, J.}, \bibinfo{author}{Hamer, C.} \&
  \bibinfo{author}{Zheng, W.}
\newblock \emph{\bibinfo{title}{Series Expansion Methods for Strongly
  Interacting Lattice Models}} (\bibinfo{publisher}{Cambridge University
  Press}, \bibinfo{address}{Cambridge, UK}, \bibinfo{year}{2010}).

\bibitem{Manmana2011}
\bibinfo{author}{Manmana, S.~R.}, \bibinfo{author}{Hazzard, K. R.~A.},
  \bibinfo{author}{Chen, G.}, \bibinfo{author}{Feiguin, A.~E.} \&
  \bibinfo{author}{Rey, A.~M.}
\newblock \bibinfo{title}{SU$(N)$ magnetism in chains of ultracold alkaline-earth-metal atoms: Mott transitions and quantum correlation}.
\newblock \emph{\bibinfo{journal}{Phys. Rev. A}} \textbf{\bibinfo{volume}{84}},
  \bibinfo{pages}{043601} (\bibinfo{year}{2011}).

\bibitem{Messio2012}
\bibinfo{author}{{Messio}, L.} \& \bibinfo{author}{{Mila}, F.}
\newblock \bibinfo{title}{Entropy dependence of correlations in one-dimensional $\mathrm{SU}(N)$ antiferromagnets}.
\newblock \emph{\bibinfo{journal}{Phys. Rev. Lett.}}
  \textbf{\bibinfo{volume}{109}}, \bibinfo{pages}{205306}
  (\bibinfo{year}{2012}).

\bibitem{Capponi2015}
\bibinfo{author}{{Capponi}, S.}, \bibinfo{author}{{Lecheminant}, P.} \&
  \bibinfo{author}{{Totsuka}, K.}
\newblock \bibinfo{title}{{Phases of one-dimensional SU($N$) cold atomic Fermi
  gases --from molecular Luttinger liquids to topological phases}}.
\newblock \eprint{arXiv:1509.04597} (\bibinfo{year}{2015}).

\bibitem{Hung2011}
\bibinfo{author}{Hung, C.-L.} \emph{et~al.}
\newblock \bibinfo{title}{Extracting density–density correlations from in
  situ images of atomic quantum gases}.
\newblock \emph{\bibinfo{journal}{New J. Phys.}} \textbf{\bibinfo{volume}{13}},
  \bibinfo{pages}{075019} (\bibinfo{year}{2011}).

\end{thebibliography}

\begin{thebibliography}{3}
\expandafter\ifx\csname url\endcsname\relax
  \def\url#1{\texttt{#1}}\fi
\expandafter\ifx\csname urlprefix\endcsname\relax\def\urlprefix{URL }\fi
\providecommand{\bibinfo}[2]{#2}
\providecommand{\eprint}[2][]{\url{#2}}

\bibitem[S1]{reinaudi_strong_2007a}
\bibinfo{author}{Reinaudi, G.}, \bibinfo{author}{Lahaye, T.},
  \bibinfo{author}{Wang, Z.} \& \bibinfo{author}{Guéry-Odelin, D.}
\newblock \bibinfo{title}{Strong saturation absorption imaging of dense clouds
  of ultracold atoms}.
\newblock \emph{\bibinfo{journal}{Opt. Lett.}}
  \textbf{\bibinfo{volume}{32}}, \bibinfo{pages}{3143} (\bibinfo{year}{2007}).


\bibitem[S2]{esry_recombination_1999}
\bibinfo{author}{Esry, B.~D.}, \bibinfo{author}{Greene, C.~H.} \&
  \bibinfo{author}{Burke, J.~P.}
\newblock \bibinfo{title}{Recombination of three atoms in the ultracold limit}.
\newblock \emph{\bibinfo{journal}{Phys. Rev. Lett.}}
  \textbf{\bibinfo{volume}{83}}, \bibinfo{pages}{1751--1754}
  (\bibinfo{year}{1999}).

\bibitem[S3]{jack_signatures_2003}
\bibinfo{author}{Jack, M.~W.} \& \bibinfo{author}{Yamashita, M.}
\newblock \bibinfo{title}{Signatures of the quantum fluctuations of cold atoms
  in an optical lattice in the three-body loss rate}.
\newblock \emph{\bibinfo{journal}{Phys. Rev. A}}
  \textbf{\bibinfo{volume}{67}}, \bibinfo{pages}{033605}
  (\bibinfo{year}{2003}).

\end{thebibliography}


\renewcommand{\thefigure}{S\arabic{figure}}
 \setcounter{figure}{0}
\renewcommand{\theequation}{S.\arabic{equation}}
 \setcounter{equation}{0}
 \renewcommand{\thesection}{S.\Roman{section}}
\setcounter{section}{0}
\renewcommand{\thetable}{S\arabic{table}}
 \setcounter{table}{0}


\onecolumngrid

\clearpage

\begin{center}
	\noindent\textbf{Supplementary Information:}
	\bigskip
	
	\noindent\textbf{\large{Direct probing of the Mott crossover in the SU(\textit{N}) Fermi-Hubbard model}}
	
	\bigskip
	Christian Hofrichter, Luis Riegger, Francesco Scazza, Moritz H\"ofer,\\ \mbox{Diogo Rio Fernandes}, Immanuel Bloch, and Simon F\"olling
	\vspace{0.1cm}
	
	\small{ \emph{Fakult\"at f\"ur Physik, Ludwig-Maximilians-Universit\"at, Schellingstrasse 4, 80799 M\"unchen, Germany and}}
	
	\small{ \emph{Max-Planck-Institut f\"ur Quantenoptik, Hans-Kopfermann-Strasse 1, 85748 Garching, Germany}}
\end{center}
\bigskip

\section{High-intensity imaging calibration}

In order to correctly determine the density of the gas using high-intensity absorption imaging, we use the modified Lambert-Beer law \cite{reinaudi_strong_2007a} which accounts for saturation of the imaging transition:

\begin{equation}
OD = -\alpha \ln\left(\frac{I_f}{I_i}\right) + \frac{I_i - I_f}{I_\mathrm{sat}},
\label{eq:mod_lambertbeer}
\end{equation}
where $OD$ is the optical density, $I_\mathrm{i}$ the incident light intensity, $I_\mathrm{f}$ the final light intensity after absorption and $I_\mathrm{sat}$ the saturation intensity. The saturation intensity $I_\mathrm{sat} \simeq 60\mathrm{mW}/\mathrm{cm}^2$ is calculated from the linewidth and wavelength of the transition, and additionally verified by an intensity-dependent linewidth measurement.

The $\alpha$-parameter is extracted by varying the saturation parameter $I/I_\mathrm{sat}$ and adjusting its value in a way that the measured optical density from equation \eqref{eq:mod_lambertbeer} is independent of the light intensity  \cite{reinaudi_strong_2007a}. It accounts for a transition strength lower than 1 and, potentially, a systematic error when determining the value of the light intensity impinging on the atoms.

\begin{figure}[htpb]
\centering
\includegraphics[width=1\columnwidth]{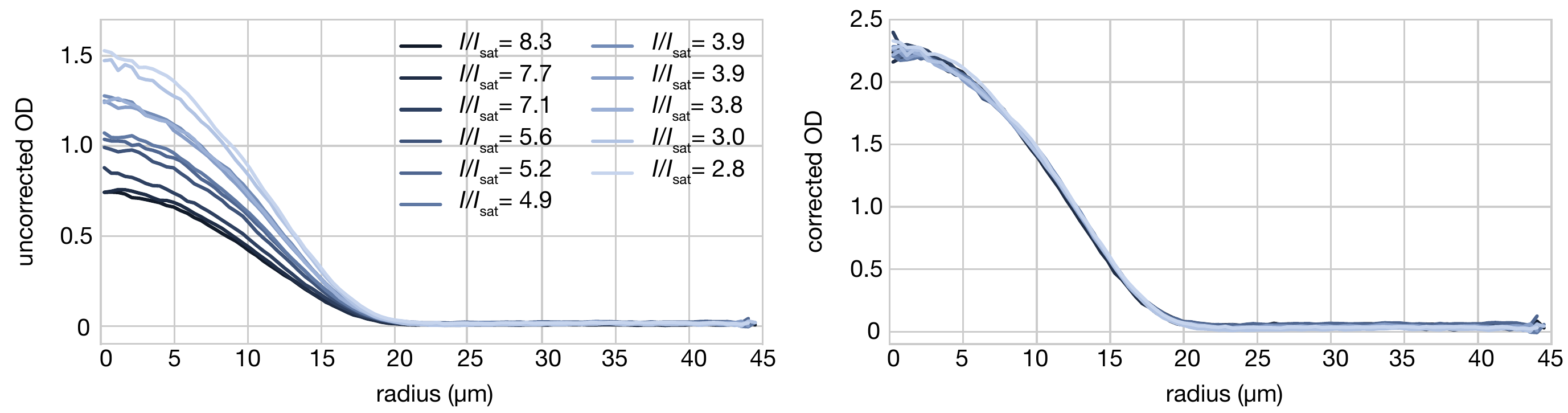}
\caption{Density profiles of the harmonically trapped SU($6$) gas for various saturation parameters without  saturation correction (left) and with  saturation correction of Eq. \eqref{eq:mod_lambertbeer} with $\alpha = 3.05$ (right).}
\label{fig:alpha_extraction2}
\end{figure}

\newpage

\section{Characterization of three-body losses}

In order to avoid three-body losses in the lattice we keep the central average filling below $n d^3 \simeq 1.7$. For higher filling factors we observe a fast density decay in the central region of the cloud in addition to a slow density decay due to vacuum losses and technical heating of the lattice beams. We fit a double exponential decay to  the core density of the cloud
\begin{equation}
n(t)=n_d*e^{-t/\tau_v}+n_3*e^{-t(3/\tau_v+1/\tau_3)}
\label{eq:threebody_loss}
\end{equation}
where $n_d$ is the density of singly and doubly occupied sites, $n_3$ is the density of triply occupied sites and $\tau_v$,$\tau_3$ denote the vacuum lifetime and three-body loss timescale respectively.  We extract a loss rate $\gamma=1/\tau_3=2.4(3)\,$Hz in a $15E_\mathrm{r}$ deep lattice.  The scaling of the loss rate as a function of the lattice depth is found to be compatible with the expected scaling of a three-body decay rate 
\begin{equation}
\gamma = \beta_3 \int{d^{3} \mathbf{x}\, w( \mathbf{x} )^6}
\label{eq:K3coef}
\end{equation}
where $w( \mathbf{x} )$ is the lattice depth-dependent Wannier function \cite{esry_recombination_1999,jack_signatures_2003}. Using Eq. (\ref{eq:K3coef}), we determine the $\beta_3$ coefficient independently for three lattice depths $V=15\,E_\mathrm{r}$, $V=22\,E_\mathrm{r}$ and $V=30\,E_\mathrm{r}$. From these measurements, we obtain an estimate of the three-body loss rate coefficient $\beta_3=2.3(6) \times 10^{-29} \mathrm{cm^6/s}$. This value is of the same order of magnitude of previously reported three-body loss rate coefficients in alkali atoms \cite{esry_recombination_1999}. 

\begin{figure}[htpb]
\centering
\includegraphics[width=.6\columnwidth]{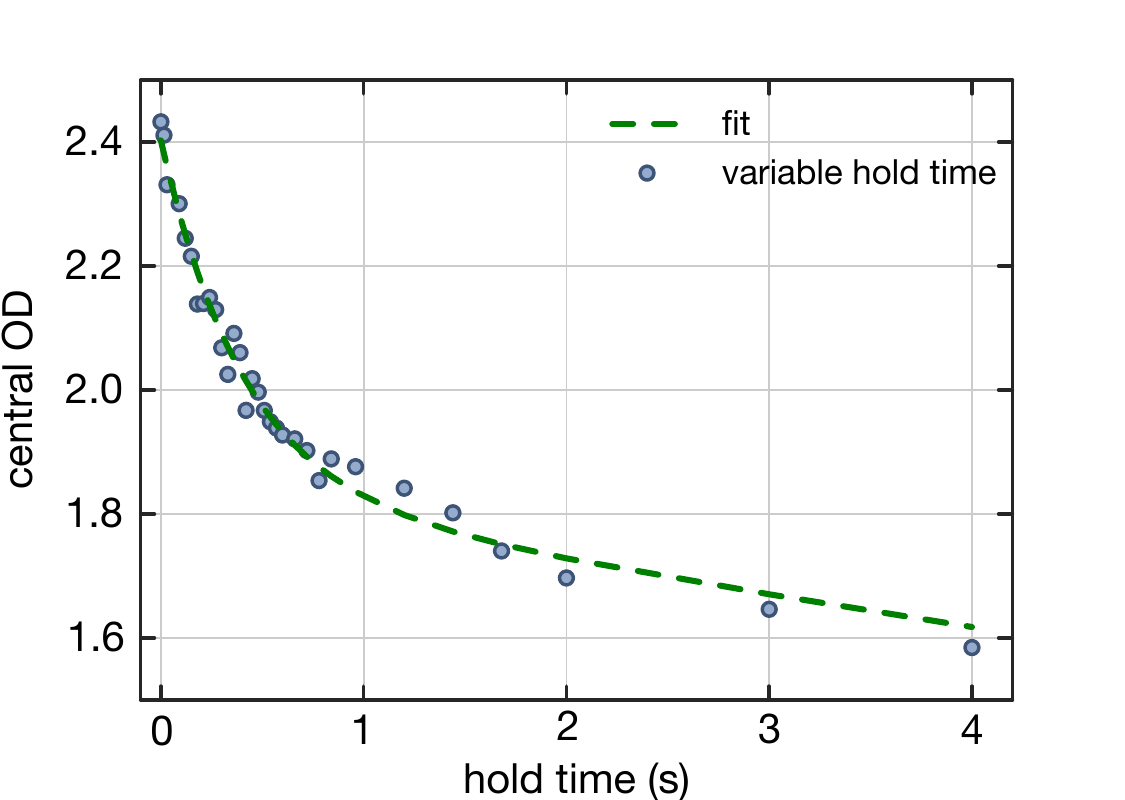}
\caption{Atom loss at high densities in the center of the cloud in $V=15E_\mathrm{r}$ deep lattice. The fit function is a double exponential decay described by Eq. \eqref{eq:threebody_loss}.  }
\label{fig:heating_lattice_loading}
\end{figure}

\end{document}